\documentclass{article}
\usepackage[psamsfonts]{amssymb}
\usepackage{amsmath}
\usepackage{cite}
\usepackage{bm}
\usepackage{graphicx}
\usepackage{yhmath}
\usepackage{bbm}

\def\d{\mathrm{d}}

\def\id{\mathbf{1}}

\def\ir{\mathrm{i}}

\begin{document}
\begin{titlepage}
\noindent{\large\textbf{Quantum mechanics on space with SU(2) fuzziness}}

\vspace{\baselineskip}

\begin{center}
{Amir~H.~Fatollahi~{\footnote {ahfatol@gmail.com}}\\
\vskip 0.1cm
Ahmad~Shariati~{\footnote {shariati@mailaps.org}}}\\
\vskip 0.1cm
Mohammad~Khorrami~{\footnote {mamwad@mailaps.org}}\\
\vskip 10 mm
\textit{ Department of Physics, Alzahra University, Tehran
1993891167, Iran }
\end{center}
\vspace{\baselineskip}
\begin{abstract}
\noindent Quantum mechanics of models is considered which are
constructed in spaces with Lie algebra type commutation relations
between spatial coordinates. The case is specialized to that of
the group SU(2), for which the formulation of the problem via the
Euler parameterization is also presented. SU(2)-invariant systems
are discussed, and the corresponding eigenvalue problem for the
Hamiltonian is reduced to an ordinary differential equation, as it
is the case with such models on commutative spaces.
\end{abstract}
\end{titlepage}

\section{Introduction}
In recent years there has been great interest to study physical
models on spaces with noncommutative coordinates. In the simplest
case of canonical noncommutative space the coordinates satisfy
\begin{equation}\label{kfs.1}
[\hat{x}_\mu,\hat{x}_\nu]=\ir\,\theta_{\mu\,\nu}\,\id,
\end{equation}
in which $\theta$ is an antisymmetric constant tensor and $\id$
represents the unit operator. It has been understood that the
longitudinal directions of D-branes in the presence of a constant
B-field background appear to be noncommutative, as seen by the
ends of open strings \cite{9908142,99-2,99-3,99-4}. The
theoretical and phenomenological implications of such
noncommutative coordinates have been extensively studied
\cite{reviewnc}.

One direction to extend studies on noncommutative spaces is to
consider spaces where the commutators of the coordinates are not
constants. Examples of this kind are the noncommutative cylinder
and the $q$-deformed plane \cite{chai}, the so-called
$\kappa$-Poincar\'{e} algebra \cite{majid,ruegg,amelino,kappa},
and linear noncommutativity of the Lie algebra type
\cite{wess,sasak}. In the latter the dimensionless spatial
positions operators satisfy the commutation relations of a Lie
algebra:
\begin{equation}\label{kfs.2}
[\hat{x}_a,\hat{x}_b]= f^c{}_{a\, b}\,\hat{x}_c,
\end{equation}
where $f^c{}_{a\,b}$'s are structure constants of a Lie algebra.
One example of this kind is the algebra SO(3), or SU(2). A special
case of this is the so called fuzzy sphere \cite{madore,presnaj},
where an irreducible representation of the position operators is
used which makes the Casimir of the algebra,
$(\hat{x}_1)^2+(\hat{x}_2)^2+(\hat{x}_3)^2$, a multiple of the
identity operator (a constant, hence the name sphere). One can
consider the square root of this Casimir as the radius of the
fuzzy sphere. This is, however, a noncommutative version of a
two-dimensional space (sphere).

In \cite{0612013,fakE1,fakE2} a model was introduced in which the
representation was not restricted to an irreducible one, instead
the whole group was employed. In particular the regular
representation of the group was considered, which contains all
representations. As a consequence in such models one is dealing
with the whole space, rather than a sub-space, like the case of
fuzzy sphere as a 2-dimensional surface. In \cite{0612013} basic
ingredients for calculus on a linear fuzzy space, as well as basic
notions for a field theory on such a space, were introduced. In
\cite{fakE1} basic elements for calculating the matrix elements
corresponding to transition between initial and final states were
discussed. Models based on the regular representation of SU(2)
were treated in more detail, giving explicit forms of the tools
and notions introduced in their general forms
\cite{0612013,fakE1}. In \cite{fakE1} and \cite{fakE2} the tree
and 1-loop diagrams for a self-interacting scalar field theory
were discussed, respectively. It is observed that models based on
Lie algebra type noncommutativity enjoy three features:
\begin{itemize}
\item They are free from any ultraviolet divergences if the group
is compact. \item There is no momentum conservation in such
theories. \item In the transition amplitudes only the so-called
planar graphs contribute.
\end{itemize}
The reason for latter is that the non-planar graphs are
proportional to $\delta$-distributions whose dimensions are less
than their analogues coming from the planar sector, and so their
contributions vanish in the infinite-volume limit usually taken in
transition amplitudes \cite{fakE2}.

The facts that in such theories the mass-shell condition is
different, and there is no momentum conservation, lead to
different consequences (with respect to ordinary theories) in
collisions. This was exploited in \cite{skf}, where it was seen
that there may be a new threshold for the collision of two
massless particles to produce massive particles.

In \cite{kfs} the classical mechanics defined on a space with
SU(2) fuzziness was studied. In particular, the Poisson structure
induced by noncommutativity of SU(2) type was investigated, for
either Cartesian or Euler parameterization of SU(2) group. The
consequences of SU(2)-symmetry in such spaces on integrability,
was also studied in \cite{kfs}.

The purpose of the present work is to examine the quantum
mechanics on a space with SU(2) fuzziness. In particular quantum
models are studied which have SU(2)-symmetry.

The scheme of the rest of this paper is the following. In section
2, the commutation relations of the position and momentum
operators corresponding to spaces with Lie-algebra
noncommutativity in the configuration space are studied. In
section 3, these are specialized to the group SU(2). In section 4
systems are studied which are SU(2)-invariant, and the eigen-value
problem for the corresponding Hamiltonian is reduced to an
ordinary differential equation.

\section{The quantum commutators}
Consider a Lie group G. Denote the members of a basis for the
left-invariant vector fields corresponding to this group by
$\hat{x}_a$'s. These fields satisfy (\ref{kfs.2}), with the
structure constants of the Lie algebra corresponding to G. The
coordinates $\hat{k}^a$ are defined such that
\begin{equation}\label{kfs.3}
U(\hat{\mathbf{k}}):=[\,\exp(\hat{k}^a\,\hat{x}_a)]\,U(\mathbf{0}),
\end{equation}
where $U(\hat{\mathbf{k}})$ is the group element corresponding to
the coordinates $\hat{\mathbf{k}}$, $U(\mathbf{0})$ is the
identity, and $\exp(\hat{x})$ is the flux corresponding to the
vector field $\hat{x}$. The action of $L_{\hat{x}_a}$ (the Lie
derivative corresponding to the vector field $\hat{x}_a$) on an
arbitrary scalar function $F$ can be written like
\begin{equation}\label{kfs.4}
L_{\hat{x}_a}(F)=\hat{x}_a{}^b\,\frac{\partial\,F}{\partial\hat{k}^b},
\end{equation}
where $\hat{x}_a{}^b$'s are scalar functions, and satisfy
\begin{equation}\label{kfs.5}
\hat{x}_a{}^b(\hat{\mathbf{k}}=\mathbf{0})=\delta_a^b.
\end{equation}
One can define the vector fields $\hat{X}_a$ locally through
\begin{equation}\label{kfs.6}
L_{\hat{X}_a}(F)=\frac{\partial\,F}{\partial\hat{k}^a},
\end{equation}
so that
\begin{equation}\label{kfs.7}
\hat{x}_a=\hat{x}_a{}^b\,\hat{X}_b.
\end{equation}
Then, considering scalar functions as operators acting on scalar
functions through simple multiplications, and vector fields as
operators acting on scalar functions through Lie derivation, one
arrives at the following commutation relations.
\begin{align}\label{kfs.8}
[\hat{X}_a,\hat{X}_b]&=0,\\ \label{kfs.9}
[\hat{X}_a,\hat{k}^b]&=\delta_a^b,\\ \label{kfs.10}
[\hat{k}^a,\hat{k}^b]&=0.
\end{align}
One should, however, remember that the functions $\hat{k}^a$ and
the vector fields $\hat{X}_a$ are only locally defined. One can
write the above commutation relations in terms of $\hat{x}_a$'s
instead of $\hat{X}_a$'s. The equation corresponding to
(\ref{kfs.8}) would be (\ref{kfs.2}), while that corresponding to
(\ref{kfs.9}) would be
\begin{equation}\label{kfs.11}
[\hat{x}_a,\hat{k}^b]=\hat{x}_a{}^b,
\end{equation}
and as $\hat{x}_a{}^b$'s are scalar functions, they commute with
$\hat{k}^a$'s.

Next consider the right-invariant vector fields
$\hat{x}_a^{\mathrm{R}}$, so that they coincide with their
left-invariant analogues at the identity of the group:
\begin{equation}\label{kfs.12}
\hat{x}_a^{\mathrm{R}}(\hat{\mathbf{k}}=\mathbf{0})=
\hat{x}_a(\hat{\mathbf{k}}=\mathbf{0}).
\end{equation}
These field satisfy the commutation relations
\begin{align}\label{kfs.13}
[\hat{x}_a^{\mathrm{R}},\hat{x}_b^{\mathrm{R}}]&=
-f^c{}_{a\,b}\,\hat{x}_c^{\mathrm{R}},\\ \label{kfs.14}
[\hat{x}_a^{\mathrm{R}},\hat{x}_b]&=0.
\end{align}
Using these, one defines the new vector field $\hat{J}_a$ through
\begin{equation}\label{kfs.15}
\hat{J}_a:=\hat{x}_a-\hat{x}_a^{\mathrm{R}}.
\end{equation}
These are the generators of the adjoint action, and satisfy the
commutation relations
\begin{align}\label{kfs.16}
[\hat{J}_a,\hat{J}_b]&=f^c{}_{a\,b}\,\hat{J}_c,\\
\label{kfs.17} [\hat{J}_a,\hat{X}_b]&=f^c{}_{a\,b}\,\hat{X}_c,\\
\label{kfs.18} [\hat{J}_a,\hat{x}_b]&=f^c{}_{a\,b}\,\hat{x}_c,\\
\label{kfs.19} [\hat{k}^c\,\hat{J}_a]&=f^c{}_{a\,b}\,\hat{k}^b.
\end{align}
Equations (\ref{kfs.9}), (\ref{kfs.17}), and (\ref{kfs.19}) show
that
\begin{equation}\label{kfs.20}
\hat{J}_a=-f^c{}_{a\,b}\,\hat{k}^b\,\hat{X}_c.
\end{equation}
Equation (\ref{kfs.9}) ensures that there is no ambiguity in the
order of $\hat{k}^b$ and $\hat{X}_c$ in the above.

Using the operators introduced in the above, one can easily
construct the corresponding quantum operators. All one needs is to
multiply these operators by suitable factors to make them
Hermitian with proper dimension:
\begin{align}\label{kfs.21}
p^a&:=(\hbar/\ell)\,\hat{k}^a,\\
\label{kfs.22} X_a&:=\ir\,\ell\,\hat{X}_a,\\
\label{kfs.23} x_a&:=\ir\,\ell\,\hat{x}_a,\\
\label{kfs.24} x_a{}^b(\mathbf{p})&:=
\hat{x}_a{}^b[(\ell/\hbar)\,\mathbf{p}],\\
\label{kfs.25} J_a&:=\ir\,\hbar\,\hat{J}_a,
\end{align}
where $\ell$ is a constant of dimension length. One then arrives
at the following commutation relations.
\begin{align}\label{kfs.26}
[p^a,p^b]&=0,\\
\label{kfs.27} [X_a,p^b]&=\ir\,\hbar\,\delta_a^b,\\
\label{kfs.28} [X_a,X_b]&=0,\\
\label{kfs.29} [x_a,p^b]&=\ir\,\hbar\,x_a{}^b,\\
\label{kfs.30} [x_a,x_b]&=\ir\,\ell\,f^c{}_{a\,b}\,x_c,\\
\label{kfs.31} [J_a,X_b]&=\ir\,\hbar\,f^c{}_{a\,b}\,X_c,\\
\label{kfs.32} [J_a,x_b]&=\ir\,\hbar\,f^c{}_{a\,b}\,x_c,\\
\label{kfs.33} [p^c,J_a]&=\ir\,\hbar\,f^c{}_{a\,b}\,p^b,\\
\label{kfs.34} [J_a,J_b]&=\ir\,\hbar\,f^c{}_{a\,b}\,J_c,
\end{align}
Using (\ref{kfs.5}), it is seen that in the limit $\ell\to 0$ the
ordinary commutation relations are retrieved.

\section{The group SU(2), and the Euler parameters}
For the group SU(2), one also can define the Euler parameters
through
\begin{equation}\label{kfs.35}
[\exp(\phi\,T_3)]\,[\exp(\theta\,T_2)]\,[\exp(\psi\,T_3)]:=
[\exp(\hat{k}^a\,T_a)],
\end{equation}
where $T_a$'a are the generators of SU(2) satisfying the
commutation relation
\begin{equation}\label{kfs.36}
[T_a,T_b]=\epsilon^c{}_{a\,b}\,T_c.
\end{equation}
Using these, one arrives at
\begin{align}\label{kfs.37}
L_{\hat x_1}(F)&=-\frac{\cos\psi}{\sin\theta}\,\frac{\partial
F}{\partial\phi}+\sin\psi\,\frac{\partial F}{\partial\theta}+
\frac{\cos\psi\,\cos\theta}{\sin\theta}\,\frac{\partial
F}{\partial\psi},\\ \label{kfs.38} L_{\hat
x_2}(F)&=\frac{\sin\psi}{\sin\theta}\,\frac{\partial
F}{\partial\phi}+\cos\psi\,\frac{\partial F}{\partial\theta}-
\frac{\sin\psi\,\cos\theta}{\sin\theta}\,\frac{\partial
F}{\partial\psi},\\ \label{kfs.39} L_{\hat x_3}(F)&=\frac{\partial
F}{\partial\psi},
\end{align}
and
\begin{align}\label{kfs.40}
L_{\hat
J_1}(F)=\;&\frac{\cos\phi\,\cos\theta-\cos\psi}{\sin\theta}\,\frac{\partial
F}{\partial\phi}+(\sin\phi+\sin\psi)\,\frac{\partial
F}{\partial\theta}\cr &+
\frac{-\cos\phi+\cos\psi\,\cos\theta}{\sin\theta}\,\frac{\partial
F}{\partial\psi},\\ \label{kfs.41} L_{\hat
J_2}(F)=\;&\frac{\sin\phi\,\cos\theta+\sin\psi}{\sin\theta}\,\frac{\partial
F}{\partial\phi}+(-\cos\phi+\cos\psi)\,\frac{\partial
F}{\partial\theta}\cr & +
\frac{-\sin\phi-\sin\psi\,\cos\theta}{\sin\theta}\,\frac{\partial
F}{\partial\psi},\\ \label{kfs.42} L_{\hat
J_3}(F)=\;&-\frac{\partial F}{\partial\phi}+\frac{\partial
F}{\partial\psi},
\end{align}
for an arbitrary scalar field $F$. One also has
\begin{equation}\label{kfs.43}
\cos \frac{\hat k}{2} = \cos \frac{\theta}{2}\,
\cos\frac{\phi+\psi}{2},
\end{equation}
where
\begin{equation}\label{kfs.44}
\hat k:=\sqrt{\delta_{a\,b}\,\hat{k}^a\,\hat{k}^b}.
\end{equation}
Euler parameterization is just an alternative parameterization of
$k_a$'s as the momenta. Corresponding to these, one introduces the
coordinate operators $\hat X_\phi$, $\hat X_\theta$, and $\hat
X_\psi$. These satisfy
\begin{equation}\label{kfs.45}
[\hat{X}_\alpha,\hat{k}^\beta]=\delta^\beta_\alpha,
\end{equation}
where $\alpha$ and $\beta$ are $\phi$, $\theta$, or $\psi$, and
$k^\beta$ has been defined as $\beta$ itself. All other
commutators vanish. The simplest realization for the above
coordinate operators so that these operators are anti-Hermitian as
well, is
\begin{equation}\label{kfs.46}
\hat{X}_\alpha=\frac{1}{\sqrt{|\det\nu|}}\,\partial_\alpha\,
\sqrt{|\det\nu|},
\end{equation}
where $\nu$ is the weight function appearing in the Haar measure
$\d\mu$:
\begin{equation}\label{kfs.47}
\d\mu=\nu\,\d\phi\,\d\theta\,\d\psi.
\end{equation}
Knowing that
\begin{equation}\label{kfs.48}
\nu=c\,|\sin\theta|,
\end{equation}
where $c$ is a constant, it turns out that
\begin{align}\label{kfs.49}
\hat X_\phi&=\frac{\partial}{\partial\phi},\nonumber\\
\hat X_\theta&=\frac{1}{\sqrt{|\sin\theta|}}\,
\frac{\partial}{\partial\theta}\,\sqrt{|\sin\theta|},\nonumber\\
\hat X_\psi&=\frac{\partial}{\partial\psi}.
\end{align}
Then, one can use the differential operators in the right-hand
sides of (\ref{kfs.37}) to (\ref{kfs.42}) as realizations of
$x_a$'s and $J_a$'s, provided the following changes are performed
on them. The changes are symmetrization with respect to Euler
coordinates and their corresponding differential operators, using
$\hat X_\alpha$ instead of $\partial_\alpha$, and proper scaling
so that the dimensions of the operators are correct and the
operators are Hermitian. These result in the following
realization
\begin{align}\label{kfs.50}
x_1=&\;\ir\,\ell\,\left(-\frac{\cos\psi}{\sin\theta}\,
\frac{\partial}{\partial\phi}+\sin\psi\,\frac{\partial}{\partial\theta}+
\frac{\cos\psi\,\cos\theta}{\sin\theta}\,\frac{\partial}{\partial\psi}\right),\\
\label{kfs.51}
x_2=&\;\ir\,\ell\,\left(\frac{\sin\psi}{\sin\theta}\,
\frac{\partial}{\partial\phi}+\cos\psi\,\frac{\partial}{\partial\theta}-
\frac{\sin\psi\,\cos\theta}{\sin\theta}\,\frac{\partial}{\partial\psi}\right),\\
\label{kfs.52} x_3=&\;\ir\,\ell\,\frac{\partial}{\partial\psi},\\
\label{kfs.53}
J_1=&\;\ir\,\hbar\,\left[\frac{\cos\phi\,\cos\theta-\cos\psi}{\sin\theta}\,
\frac{\partial}{\partial\phi}+(\sin\phi+\sin\psi)\,
\frac{\partial}{\partial\theta}\right.\nonumber\\
&\left.+\frac{-\cos\phi+\cos\psi\,\cos\theta}{\sin\theta}\,\frac{\partial}{\partial\psi}\right],\\
\label{kfs.54}
J_2=&\;\ir\,\hbar\,\left[\frac{\sin\phi\,\cos\theta+\sin\psi}{\sin\theta}\,
\frac{\partial}{\partial\phi}+(-\cos\phi+\cos\psi)\,
\frac{\partial}{\partial\theta}\right.\nonumber\\
&\left.+\frac{-\sin\phi-\sin\psi\,\cos\theta}{\sin\theta}\,\frac{\partial}{\partial\psi}\right],\\
\label{kfs.55}
J_3=&\;\ir\,\hbar\,\left(-\frac{\partial}{\partial\phi}+
\frac{\partial}{\partial\psi}\right),
\end{align}
Introducing the new parameters $\chi$ and $\xi$:
\begin{align}\label{kfs.56}
\chi&:=\frac{\phi-\psi}{2} \\ \label{kfs.57}
\xi&:=\frac{\phi+\psi}{2},
\end{align}
it is seen that
\begin{align}\label{kfs.58}
J_\pm&=\ir\,\hbar\,\exp(\pm\,\ir\,\chi)\, \left(-\tan
\frac{\theta}{2}\,\cos\xi\,\frac{\partial}{\partial\xi}+
2\,\sin\xi\,\frac{\partial}{\partial\theta}
\pm\ir\,\cot\frac{\theta}{2}\,\sin\xi\,\frac{\partial}{\partial
\chi} \right)\\
\label{kfs.59} J_3&=-\ir\,\hbar\,\frac{\partial}{\partial\chi},
\end{align}
where
\begin{equation}\label{kfs.60}
J_\pm=J_1\pm\ir\,J_2.
\end{equation}
Again introducing new variables
\begin{align}\label{kfs.61}
v&:=\cos\frac{\theta}{2}\,\cos\xi,\\ \label{kfs.62}
\tau&:=(1-v^2)^{-1/2}\,\cos\frac{\theta}{2}\,\sin\xi,\\
\label{kfs.63} s^2&:=1-\tau^2,
\end{align}
one arrives at
\begin{align}\label{kfs.64}
J_\pm&=\ir\,\hbar\,\exp(\pm\,\ir\,\chi)\,\left(-\sqrt{1-\tau^2}\,
\frac{\partial}{\partial\tau}\pm\,\ir\,\frac{\tau}{\sqrt{1-\tau^2}}\,
\frac{\partial}{\partial\chi}\right),\nonumber\\
&=\ir\,\hbar\,\exp(\pm\,\ir\,\chi)\,\sqrt{1-s^2}\,
\left(\frac{\partial}{\partial s}\pm\,\frac{\ir}{s}\,
\frac{\partial}{\partial\chi}\right),\\ \label{kfs.65}
J_3&=-\ir\,\hbar\,\frac{\partial}{\partial\chi},
\end{align}
resulting in
\begin{align}
\mathbf{J}\cdot\mathbf{J}&=-\hbar^2\,\left[(1-\tau^2)\,
\frac{\partial^2}{\partial\tau^2}-2\,\tau\,
\frac{\partial}{\partial\tau}+\frac{1}{1-\tau^2}\,
\frac{\partial^2}{\partial\chi^2}\right],\nonumber\\
\label{kfs.66}
&=-\hbar^2\,\left[(1-s^2)\,\frac{\partial^2}{\partial
s^2}+\frac{1-2\,s^2}{s}\, \frac{\partial}{\partial
s}+\frac{1}{s^2}\, \frac{\partial^2}{\partial\chi^2}\right],
\end{align}
where
\begin{equation}\label{kfs.67}
\mathbf{A}\cdot\mathbf{B}:=\delta_{a\,b}\,A^a\,B^b.
\end{equation}
Using (\ref{kfs.65}) and (\ref{kfs.66}), it is seen that the
angular momentum eigenfunctions ($\mathcal{Y}_l^m$'s) satisfying
\begin{align}
J_3\,\mathcal{Y}_l^m&=m\,\hbar\,\mathcal{Y}_l^m,\nonumber \\
\label{kfs.68}
\mathbf{J}\cdot\mathbf{J}\,\mathcal{Y}_l^m&=l\,(l+1)\,\hbar^2\,
\mathcal{Y}_l^m,
\end{align}
are products of an arbitrary function $f$ of $v$, and $Y_l^m$'s
(the usual spherical harmonics) with the cosine of the colatitude
equal to $\tau$ and the longitude equal to $\chi$, that is
\begin{equation}\label{kfs.69}
\mathcal{Y}_l^m=f(v)\,Y_l^m(\cos^{-1}\tau,\chi).
\end{equation}

\section{SU(2)-invariant quantum systems}
Consider a configuration space with linear SU(2)-fuzziness, and
the corresponding Hilbert space on which the momenta and
coordinates introduced in section 3 act. A system characterized by
a Hamiltonian $H$, is said to be SU(2)-invariant, if $H$ is
SU(2)-invariant, that is if the commutators of $H$ with $J_a$'s
vanish. A Hamiltonian which is a function of only
$(\mathbf{p}\cdot\mathbf{p})$ and $(\mathbf{x}\cdot\mathbf{x})$ is
clearly so. The aim is to exploit the SU(2)-symmetry of such a
Hamiltonian to write down an eigenvalue equation for the
Hamiltonian so that that equation contains only one variable (from
the three variables corresponding to the momentum). To do so, one
calculates $(\mathbf{x}\cdot\mathbf{x})$. The result is
\begin{equation}\label{kfs.70}
\mathbf{x}\cdot\mathbf{x}=-\ell^2\,\left(
\frac{1+\cos\theta}{2\,\sin^2\theta}\,
\frac{\partial^2}{\partial\chi^2}+
\frac{1-\cos\theta}{2\,\sin^2\theta}\,
\frac{\partial^2}{\partial\xi^2} +\frac{1}{\sin\theta}\,
\frac{\partial}{\partial\theta}\,\sin\theta\,
\frac{\partial}{\partial\theta}\right),
\end{equation}
or
\begin{align}\label{kfs.71}
\mathbf{x}\cdot\mathbf{x}=&\;-\ell^2\,\left\{
\frac{1}{4\,(1-v^2)}\,\left[ (1-s^2)\,\frac{\partial^2}{\partial
s^2}+\frac{1-2\,s^2}{s}\, \frac{\partial}{\partial
s}+\frac{1}{s^2}\,
\frac{\partial^2}{\partial\chi^2}\right]\right.\nonumber\\
&\left.+\frac{1-v^2}{4}\,\frac{\partial^2}{\partial
v^2}-\frac{3\,v}{4}\,\frac{\partial}{\partial v}\right\},\nonumber\\
 =&\;-\ell^2\,\left[-\frac{\hbar^{-2}}{4\,(1-v^2)}\,
\mathbf{J}\cdot\mathbf{J}+
\frac{1-v^2}{4}\,\frac{\partial^2}{\partial
v^2}-\frac{3\,v}{4}\,\frac{\partial}{\partial v}\right].
\end{align}
From this, one finds that
\begin{equation}\label{kfs.72}
\mathbf{x}\cdot\mathbf{x}\,\mathcal{Y}_l^m=\ell^2\,Y_l^m\,\left[
\frac{l\,(l+1)}{4\,(1-v^2)}- \frac{1-v^2}{4}\, \frac{\d^2}{\d
v^2}+\frac{3\,v}{4}\, \frac{\d}{\d v}\right] \,f.
\end{equation}
Substituting $f$ in (\ref{kfs.69}) with another function
$\Upsilon$,
\begin{equation}\label{kfs.73}
f(v)=:(1-v^2)^{l/2}\,\Upsilon(v),
\end{equation}
one arrives at
\begin{equation}\label{kfs.74}
\mathbf{x}\cdot\mathbf{x}\,\mathcal{Y}_l^m=\ell^2\,(1-v^2)^{l/2}\,
Y_l^m\,\left[-\frac{1-v^2}{4}\, \frac{\d^2}{\d v^2}
+\left(\frac{3}{4}+\frac{l}{2}\right)\,v\,\frac{\d}{\d v}
+\frac{l}{2}\,\left(\frac{l}{2}+1\right)\right] \,\Upsilon.
\end{equation}
Now assume that the Hamiltonian is the sum of a kinetic term $K$,
which is a function only $(\mathbf{p}\cdot\mathbf{p})$, and a
potential term $V$, which is function of only
$(\mathbf{x}\cdot\mathbf{x})$:
\begin{equation}\label{kfs.75}
H=K+V.
\end{equation}
Noting that $(\mathbf{p}\cdot\mathbf{p})$ is a function of only
$v$, the eigenvalue equation for $H$ becomes
\begin{equation}\label{kfs.76}
K\,\Upsilon+V\left\{\mathbf{x}\cdot\mathbf{x}=
\ell^2\,\left[-\frac{1-v^2}{4}\, \frac{\d^2}{\d v^2}
+\left(\frac{3}{4}+\frac{l}{2}\right)\,v\,\frac{\d}{\d v}
+\frac{l}{2}\,\left(\frac{l}{2}+1\right)\right]\right\}\,\Upsilon=E\,\Upsilon,
\end{equation}
where $K$ is a function of only $v$. An example for $K$ is \cite{fakE1,fakE2,skf,kfs}
\begin{align}\label{kfs.77}
K&=\frac{4\,\hbar^2}{\ell^2\,m}\,
\left(1-\cos\frac{\ell\,p}{2\,\hbar}\right),\nonumber\\
&=\frac{4\,\hbar^2}{\ell^2\,m}\,(1-v).
\end{align}
One then arrives at
\begin{align}\label{kfs.78}
&\frac{4\,\hbar^2}{\ell^2\,m}\,(1-v)\,\Upsilon\nonumber\\
+&V\left\{\mathbf{x}\cdot\mathbf{x}=
\ell^2\,\left[-\frac{1-v^2}{4}\, \frac{\d^2}{\d v^2}
+\left(\frac{3}{4}+\frac{l}{2}\right)\,v\,\frac{\d}{\d v}
+\frac{l}{2}\,\left(\frac{l}{2}+1\right)\right]\right\}\,\Upsilon=E\,\Upsilon,
\end{align}
If the potential function $V$ is bounded from above, the
eigenvalues of the Hamiltonian would be bounded from above, with
the following as an upper bound
\begin{equation}\label{kfs.79}
E\leq V_{\mathrm{max}}+\frac{8\,\hbar^2}{\ell^2\,m},
\end{equation}
where $V_{\mathrm{max}}$ is the maximum of $V$. If it is possible
that $V$ takes large values (compared to the maximum of $K$), then
large eigenvalues are possible for the Hamiltonian and these
correspond to eigenvectors which are approximately eigenvectors of
$(\mathbf{x}\cdot\mathbf{x})$:
\begin{align}\label{kfs.80}
(\mathbf{x}\cdot\mathbf{x})\,\Psi&=\ell^2\,j\,(j+1)\,\Psi,\nonumber\\
H\,\Psi&=E\,\Psi,\nonumber\\
\frac{E}{V[\mathbf{x}\cdot\mathbf{x}=\ell^2\,j\,(j+1)]}&\approx 1,
\end{align}
where $(2\,j)$ is a nonnegative integer.

As examples consider the free particle and the harmonic
oscillator. For the former there is no potential term, and the
spectrum of the Hamiltonian is bounded from above (and of course
below):
\begin{equation}\label{kfs.81}
0\leq E\leq\frac{8\,\hbar^2}{\ell^2\,m}.
\end{equation}
For the latter, one uses the potential
\begin{equation}\label{kfs.82}
V=\frac{1}{2}\,m\,\omega^2\,\mathbf{x}\cdot\mathbf{x},
\end{equation}
to arrive at the following equation for the eigenvalue problem.
\begin{align}\label{kfs.83}
&\left\{\frac{4\,\hbar^2}{\ell^2\,m}\,(1-v)\right.\nonumber\\
&\;\left.+\frac{\ell^2}{2}\,m\,\omega^2\, \left[-\frac{1-v^2}{4}\,
\frac{\d^2}{\d v^2}
+\left(\frac{3}{4}+\frac{l}{2}\right)\,v\,\frac{\d}{\d v}
+\frac{l}{2}\,\left(\frac{l}{2}+1\right)\right]\right\}\,\Upsilon=E\,\Upsilon.
\end{align}
As the corresponding potential is not bounded from above for large
values of $(\mathbf{x}\cdot\mathbf{x})$, one can use the above
argument to see that for large values of $j$, the eigenvectors of
$(\mathbf{x}\cdot\mathbf{x})$ corresponding to the eigenvalue
$\ell^2\,j\,(j+1)$ are also eigenvalues of the Hamiltonian, and
the corresponding energies satisfy
\begin{equation}\label{kfs.84}
\lim_{j\to\infty}\frac{E_j}{j\,(j+1)}=\frac{1}{2}\,m\omega^2\,\ell^2.
\end{equation}

\noindent\textbf{Acknowledgement}:  This work was partially
supported by the research council of the Alzahra University.

\end{document}